\documentclass[11pt,rotating]{article}
\usepackage{fleqn,cospar}

\usepackage{graphicx}
\usepackage[figuresright]{rotating}

\title{DID EGRET DETECT DISTANT SUPERNOVA REMNANTS?}

\author{Diego F. Torres\address{Lawrence Livermore Laboratory, 7000 East
Ave. L-413, Livermore, CA 94550, USA}, Gustavo E.
Romero\address{Instituto Argentino de Radioastronom\'{\i}a, C.C.5,
1894 Villa Elisa, Buenos Aires, Argentina}, Thomas M.
Dame\address{Harvard-Smithsonian Center for Astrophysics, 60
Garden Street, Cambridge, MA 02138, USA}, Jorge A. Combi$^3$,
Yousaf M. Butt$^4$}

\begin{document}

\maketitle

\begin{abstract}
It might be thought that supernova remnants (SNRs) more distant
than a few kiloparsec from Earth could not have been detected by
the EGRET experiment. This work analyzes the observational status
of this statement in the light of new CO studies of SNRs.
\end{abstract}

\section{ INTRODUCTION}

When analyzing the $\gamma$-ray detections in the COS-B and
partial EGRET catalogs, it was commonly believed that SNRs more
distant than a few kpc from Earth could not have been detected by
these experiments. The inverse square dependence of the flux on
the distance to the source was, of course, critical in reaching
such a conclusion. However, it is not only the distance but also
the molecular environment that is crucial in determining
detectability. In this presentation we analyze, in the light of
new CO studies of SNRs, whether EGRET could have detected SNRs
lying farther away than 5 kpc from Earth.

\section{ EGRET SOURCE -- SUPERNOVA REMNANT PAIRS}

\begin{sidewaystable}
\begin{center}
\caption{Positional coincidences between supernova remnants quoted
in Green's Catalog (2000), and unidentified 3EG EGRET sources. See
text for the meaning of the different columns. The distance $d$ is
in kpc. $F_{\rm radio}^{1{\rm GHz}}$ is in Jy, and $L_\gamma$ is
in erg s$^{-1}$. The distances mentioned are reported or discussed
in the references quoted below. Radio fluxes and spectral indices
($\alpha$, such that $S_{\nu}\propto\nu^{\alpha}$) are taken from
Green's (2000) Catalog. }\vspace{0.2cm} {\small
\begin{tabular}{lllllllllllllllll}
\hline $\gamma$-source & $F_{\gamma} $ & $\Gamma$ & Class  &$I$&
$\tau $& P? & ${\rm SNR}$ & Other name & $\Delta\theta$ & Size & T
& $d$ & Ref. & $F_{\rm radio}^{1{\rm GHz}}$ & $\alpha$ &
$L_\gamma$ \\

 \noalign{\smallskip} \hline \noalign{\smallskip}

0542$+$2610 & 14.7$\pm$3.2 & 2.67$\pm$0.22 & em C& 3.16 &
$0.70_{0.34}^{1.40}$& &G180.0$-$1.7& &
2.04& 180 & S & 0.8-1.6 & 1 & 65 & varies  & 9.65 10$^{33}$\\

0617$+$2238$^{1,2}$ & 51.4$\pm$3.5 & 2.01$\pm$0.06 &  C& 1.68&
$0.26_{0.15}^{0.38}$& &G189.1$+$3.0& IC443 &
0.11& 45 & S  & 1.5 & 2& 160 & 0.36 & 1.01 10$^{35}$\\

0631$+$0642$^{1,3}$  & 14.3$\pm$3.4 & 2.06$\pm$0.15 & C& 1.52
&$75.8_{7.89}^{\infty}$& &G205.5$+$0.5& Monoceros & 1.97& 220 &  S
& 0.8-1.6 &3 & 160 & 0.5  & 1.70 10$^{34}$\\

 0634$+$0521 & 15.0$\pm$3.5 & 2.03$\pm$0.26 & em C& 1.02
&$72.0_{5.15}^{\infty}$& &G205.5$+$0.5 & Monoceros& 2.03& 220 & S
&  0.8-1.6 & 3 & 160 & 0.5 &  1.85 10$^{34}$\\

1013$-$5915 & 33.4$\pm$6.0 & 2.32$\pm$0.13 & em C& 1.63
&$0.22_{0.00}^{0.46}$& y &G284.3$-$1.8&MSH 10-53 & 0.65& 24 &   S
& 2.9 & 4 & 11 & 0.3?  & 1.71 10$^{35}$ \\

1102$-$6103 & 32.5$\pm$6.2 & 2.47$\pm$0.21 & C & 1.86
&$0.00_{0.00}^{0.90}$&&G290.1$-$0.8 &MSH 11-61A& 0.12& 19 & S
& 7 & 5 & 42 & 0.4 &8.46 10$^{35}$  \\

& & &&  && &G289.7$-$0.3 & &0.75& 18 &   S & 8.2 & * & 6.2 & 0.2?  & 1.11 10$^{36}$\\

1410$-$6147$^4$  & 64.2$\pm$8.8 & 2.12$\pm$0.14 & C& 1.22
&$0.33_{0.16}^{0.55}$&y&G312.4$-$0.4&
          & 0.23& 38 &  S  & 1.9-3.1& 6 & 45 & 0.26 & 3.06 10$^{35}$\\

1639$-$4702 & 53.2$\pm$8.7 & 2.50$\pm$0.18  & em C& 1.95
&$0.00_{0.00}^{0.38}$&y&G337.8$-$0.1 &Kes 41&
0.07& 9  & S   &12.3 & 7 & 18 & 0.5 &4.17 10$^{36}$\\
       &  &  &            &   &           &   & G338.1$+$0.4  &       & 0.65& 15
&  S     & 9.9& * & 4?  &0.4  & 2.70 10$^{36}$\\

        &  & &             &  &            &  & G338.3$+$0.0   &      & 0.57& 8
&  S  & 8.6 & * & 7? & ?  &2.04 10$^{36}$ \\

1714$-$3857 & 43.6$\pm$6.5 & 2.30$\pm$0.20  & em C& 2.17 &
$0.15_{0.00}^{0.38}$&y&G348.5$+$0.0 && 0.47& 10 &  S  & 11.3 & 8 &  9  &0.4? & 3.47 10$^{36}$\\
         &  &&          &     &             & & G348.5$+$0.1    & CTB
37A    & 0.50&  15
&  S & 11.3 & 8 &  9  &0.4?  & 3.47 10$^{36}$\\
         &&  &         &    &             & & G347.3$-$0.5       &  & 0.85&  65
&  S & 6.3& 9 & ? & ?  & 1.07 10$^{36}$\\

1734$-$3232$^5$  & 40.3$\pm$6.7 &     --      & C& 2.90&
$0.00_{0.00}^{0.24}$&&G355.6$+$0.0&
& 0.16& 8  & S  & 12.6 & *  & 3? & ?  & --\\

1744$-$3011 & 63.9$\pm$7.1 & 2.17$\pm$0.08 & C& 1.80 &
$0.38_{0.20}^{0.62}$&&G359.0$-$0.9& & 0.41& 23 & S & 6 & 10 & 23 &
0.5 & 1.65 10$^{36}$\\
           && &        &     &               & & G359.1$-$0.5 &
& 0.25&  24 &  S  & 8.5-9.2 &  8-11 &  14  & 0.4? &
3.56 10$^{36}$\\

1746$-$2851$^6$ &119.9$\pm$7.4 & 1.70$\pm$0.07 & em C& 2.00 &
$0.50_{0.36}^{0.69}$&&G0.0$+$0.0 && 0.12& 3.5   & S & 8.5& 8   & 100? & 0.8?  &1.20 10$^{37}$ \\
           &&  &       &     &              &  & G0.3$+$0.0&           & 0.19&
           16
&  S    & 8.5& 12 & 22 & 0.6 &  1.20 10$^{37}$\\

1800$-$2338$^{1,7}$  & 61.3$\pm$6.7 & 2.10$\pm$0.10 & C& 1.60 &
$0.03_{0.00}^{0.32}$&y&G6.4$-$0.1 & W28& 0.17&  42 & C   & 1.6--4.2&  13 & 310   &varies & 4.04 10$^{35}$\\

1824$-$1514 & 35.2$\pm$6.5 & 2.19$\pm$0.18 & C& 3.00 &
$0.00_{0.00}^{0.51}$&y&G16.8$-$1.1 && 0.43&  30 & -- & 1.48 & ** & 2? & ? & 5.42 10$^{34}$\\

1837$-$0423 & $<$19.1         & 2.71$\pm$0.44 & C& 5.41
&$12.0_{2.17}^{\infty}$&y&G27.8$+$0.6 &&
0.58& 50    & F    & 2 & 14& 30 & varies  & 3.40 10$^{34}$\\

1856$+$0114$^8$  & 67.5$\pm$8.6 & 1.93$\pm$0.10 & em C& 2.92
&$0.80_{0.50}^{1.51}$&y&G34.7$-$0.4 &W44&
0.17& 35 & S  & 2.5 & 15 & 230 & 0.30 & 4.14 10$^{35}$\\

1903$+$0550$^4$  & 62.1$\pm$8.9 & 2.38$\pm$0.17 &  em C&2.28 &
$0.35_{0.18}^{0.60}$&y&G39.2$-$0.3 &3C396, HC24& 0.41& 8    &  S &
7.7-9.6 & 15    & 18  & 0.6 &2.12
10$^{36}$\\

2016$+$3657  & 34.7$\pm$5.7 & 2.09$\pm$0.11  & C& 2.06 &
$0.37_{0.08}^{0.75}$&&G74.9$+$1.2  &CTB 87        & 0.26& 8   &  F
& 10& 15   &  9 & varies
&2.75 10$^{36}$\\

2020$+$4017$^{1,9}$   &123.7$\pm$6.7 & 2.08$\pm$0.04 & C& 1.12
&$0.07_{0.00}^{0.18}$&?&G78.2$+$2.1& W66, $\gamma$-Cygni
   & 0.15&  60 &  S  & 1.7 &  16 &  340  & 0.5 &  3.20 10$^{35}$\\
\noalign{\smallskip} \hline \noalign{\smallskip}
\end{tabular}
} \label{green} \mbox{}\\ \end{center}{\small $^1$ Association
proposed by Sturner \& Dermer (1995) and Esposito et al. (1996).
$^2$ GeV J0617$+$2237 $^3$ GeV J0633$+$0645. $^4$ Association
proposed by Sturner \& Dermer (1995). $^5$ GeV J1732$-$3130. $^6$
GeV J1746$-$2854. $^7$ GeV J1800$-$2328. $^8$ GeV J1856$-$0115.
$^9$ GeV J2020$+$4023. GeV sources compiled in the GeV ASCA
Catalog (Roberts et al. 2001a). References quoted for the SNR
distance (Column Ref.) are 1. Anderson et al. (1996) 2. Fesen
(1984) 3. Jaffe et al. (1997) and Hensberge et al. (2000) 4. Ruiz
\& May (1986) 5. Kaspi et al. (1997) 6. Caswell \& Barnes (1985),
Case \& Bhattacharya (1999) 7. Koralesky et al. (1998) 8. Green et
al. (1997), see also Reynoso \& Mangum (2000) 9. Slane et al.
(1999) 10. Bamba et al. (2000) 11. Uchida, Morris \& Yusef-Zadeh
(1992) 12. Kassim \& Frail (1996) 13. Frail et al. (1993) and
Clark \& Caswell (1976) 14. Reich, Furst \& Sofue (1984) 15. Green
(2000) and Caswell et al. (1975) 16. Lozinskaya et al. (2000)  *
From the $\Sigma-D$ relationship presented by Case \& Bhattacharya
(1998) ** Distance assumed equal to a coincident OB association,
Romero, Benaglia \& Torres (1999). Please see Torres et al. (2002)
for full bibliographic details.}
\end{sidewaystable}

Table \ref{green} shows those 3EG sources that are spatially
coincident with SNRs listed in the latest version of Green's
Catalog (2000). From left to right, columns are for the
$\gamma$-ray source name, the measured flux in the summed EGRET
phases P1234 (in units of $10^{-8}$ ph cm$^{-2}$ s$^{-1}$), the
photon spectral index $\Gamma$, the EGRET class of source (em for
possibly extended and C for confused), the variability indices $I$
(as in Torres et al. 2001) and $\tau_{{\rm lower\; limit}}^{{\rm
upper\; limit}}$ (as in Tompkins 1999), information about
coincidences with known radio pulsars (``y'' stands for a known
pulsar within the error box), the SNR identification (including
other usual names when available), the angular distance between
the center of the $\gamma$-ray source position and the center of
the remnant (in degrees), the size of the remnant (in arcmin), and
the SNR type T (S for shell-like emission, F for filled-centre or
plerionic remnant, and C for composite). The Poisson probability
for the 19 coincidences shown in that table to be a chance effect
(computed using thousands of simulated sets of EGRET sources, by
means of a numerical code described elsewhere, Sigl et al. 2001)
is less than $1.05 \times 10^{-5}$. Table \ref{green} presents as
well other features of the SNRs in the coincident pairs. Distances
are only approximate since several different values for the same
SNR can be found in the literature. When no direct determination
is available, estimates can be made using the radio surface
brightness-to-diameter relationship, known as $\Sigma -D$ (e.g.
Case \& Bhattacharya 1998).

\begin{figure}[t] \vspace{-1cm}
\begin{center}
\includegraphics[width=8cm,height=10cm]{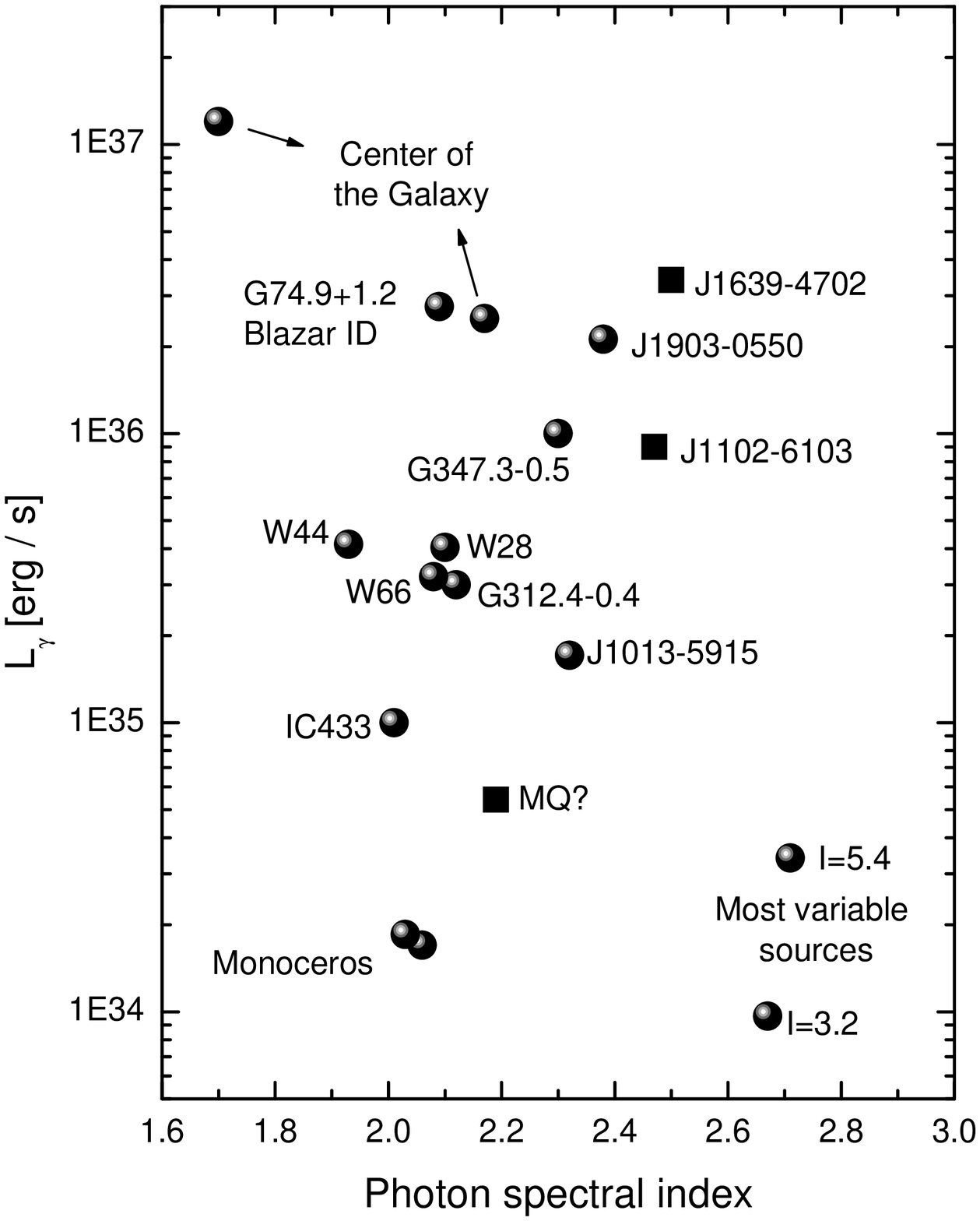}
\includegraphics[width=8cm,height=10cm]{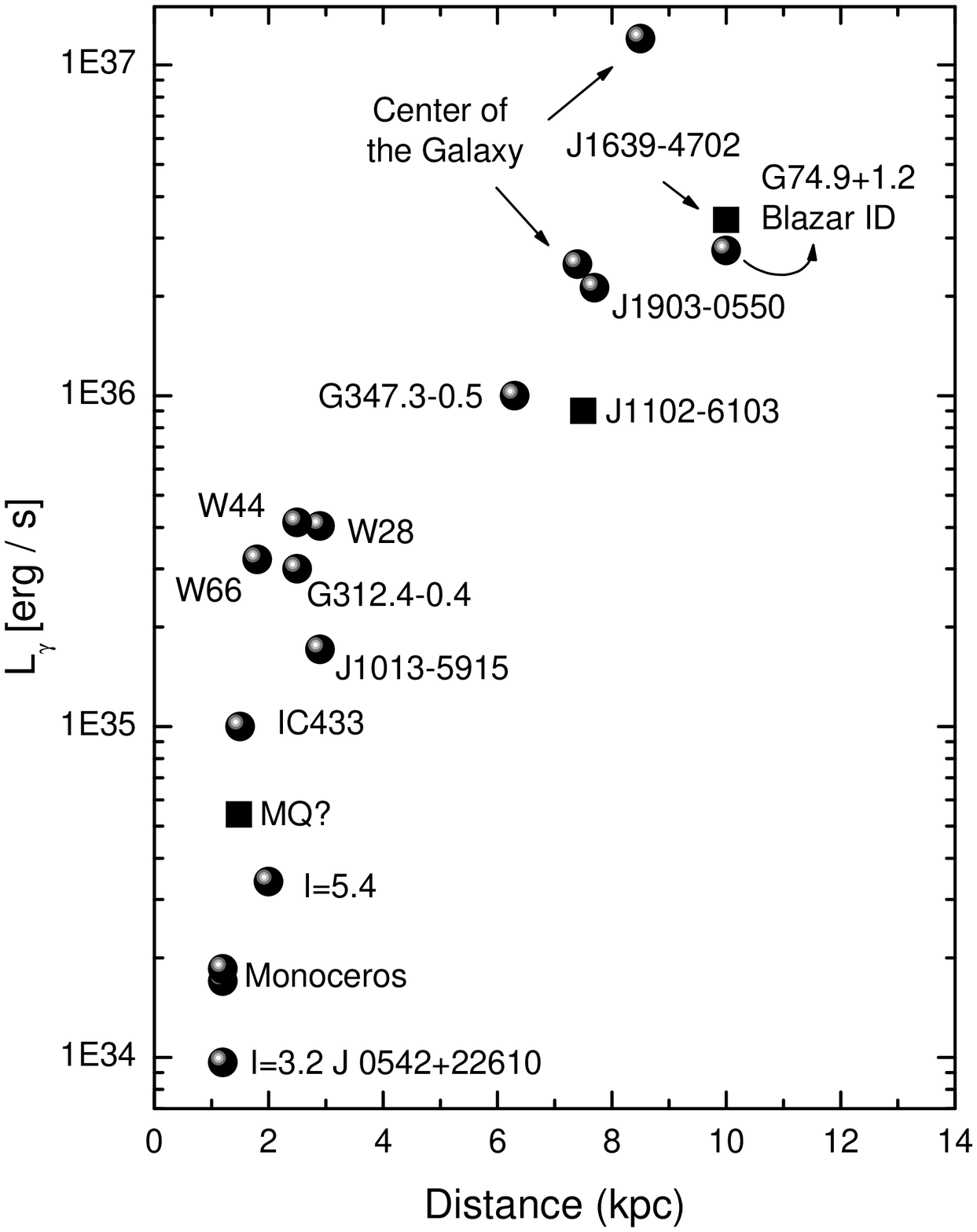}
\end{center}
\vspace{-1cm} \caption{Computed luminosities (see text) as a
function of the photon spectral index and distances. Circles stand
for SNRs whose distances has been determined, whereas squares
stand for those which distances have been estimated using the
$\Sigma-D$ relationship. } \label{lum}
\end{figure}

Using the estimated distance to each remnant in Table \ref{green},
we have calculated the approximate intrinsic $\gamma$-ray
luminosity (in the energy range 100 MeV--10 GeV, using the
observed EGRET flux and photon indices, assuming isotropic
emission). Figure \ref{lum} shows the computed luminosities as a
function of the photon spectral index and the distance to each of
the SNRs involved in Table \ref{green}.  Looking at the left panel
of Figure \ref{lum}, three different parts of that plot appear to
be distinguishable. First, at the bottom right, we find two
sources presenting the highest spectral indices together with the
highest levels of variability. It is unlikely that these two
sources are related with the involved SNRs. Three sources, at the
upper left corner, also appear, {\it a priori}, to be unlikely
related with the SNR: a tentative blazar classification has been
made for one of them (Halpern et al. 2001a), whereas for the
others, a high luminosity at too low spectral index  would be
required for Fermi's mechanism to be operative. Confusion related
to their close position towards the center of the galaxy would
make any identification even more dubious. The third region in the
plot, the central one, which apparently shows a tendency to
increase the required luminosity when increasing the photon
spectral index, encompasses the rest of the sources. All the
likely a priori candidates studied in the literature before lie
there, including the well studied W44, IC433, W28, and W66.

\section{ DISTANT SNRS AS GAMMA-RAY SOURCES?}

Disregarding the EGRET sources spatially close to, maybe even
physically associated with, the galactic center, there are 5 cases
in Table \ref{green} which coincide with SNRs apparently located
farther away than 5 kpc. Of these, 3EG J1714-3857, has been
thoroughly studied in recent years (e.g. Slane et al. 1999, Butt
et al. 2001, 2002, Muraishi et al. 2000). This is probably the
most compelling case for  acceleration of protons up to TeV
energies in a SNR, with the bulk of the observed GeV $\gamma$-rays
being produced by proton-proton interaction and subsequent pion
decays. The distance to the associated SNR, G347.3-0.5, is
estimated to be $\sim 6$ kpc (Slane et al. 1999).

In the case of 3EG J1714-3857, we associated the unidentified
EGRET GeV $\gamma$-ray source with a very massive ($\sim 3 \times
10^5 M_\odot$) and dense ($\sim 500$ nucleons cm$^{-3}$ )
molecular cloud interacting with the TeV $\gamma$-ray emitting,
spatially coincident, SNR G347.3-0.5. Since the cloud region is of
low radio and X-ray brightness, we were able to dismiss an
electronic origin for the bulk of the GeV emission there.
Furthermore, the ambient cloud medium is directly measured to be
unusually excited, suggesting that it is indeed being overtaken by
the SNR shock front. A picture then emerges where, even when the
SNR lie farther than $\sim $5 kpc, it was possible for EGRET to
detect it (Butt et al. 2001). A question then naturally arises:
could EGRET have detected even farther SNRs?

\subsection{ 3EG J1102-6103 -- SNR G290.1-0.8
(MSH 11-61A)/289.7-0.3}

The line of sight to 3EG 1102-6103 intersects both the near and
far sides of the Carina spiral arm, at velocities near $-20$ km
s$^{-1}$ and $+20$ km s$^{-1}$ respectively. There is a distinct
gap in the near side of the Carina arm in the direction of the 3EG
source, with almost no molecular gas within $\sim 1$ deg of the
source direction (see, e.g., Figure 2 of Dame, Hartmann, \&
Thaddeus 2001). On the other hand, there is a very massive
molecular complex in the far Carina Arm overlapping the direction
of the 3EG source; this complex is No. 13 in the Carina Arm cloud
catalog of Grabelsky et al. (1988). There is little doubt that the
two component clouds, labeled A and B in Figure 14 of Torres et
al. (2002) are part of the same complex, since they have
approximately the same velocity of 22 km s$^{-1}$, and are
connected smoothly by weaker emission, also at the same velocity.
Also, the HII regions are evidence of abundant on-going star
formation in this molecular complex which additionally supports
the association of the SNR. Assuming a flat rotation curve beyond
the solar circle, the kinematic distance of the complex is 8.0 kpc
and its total molecular mass is 2.1 $\times 10^6 M_\odot$.

It is worth noting that the composite CO line profile of cloud B
is very broad and complex, suggesting possible interaction with
SNR G290.1-0.8. In the case of cloud A, its radius ($\sim $48 pc)
and composite linewidth (17 km s$^{-1}$ FWHM) are roughly
consistent with the radius-linewidth relation found for large
molecular complexes by Dame et al. (1986). For cloud B, however,
its linewidth ($\sim $27 km s$^{-1}$) is about a factor of 3 too
large compared to its radius ($\sim $28 pc). The coinciding SNR
G289.7-0.3 is far from Cloud B, in a region of low molecular
density. It is extremely unlikely that this SNR is related with
the 3EG source in question. The only remaining candidate is, then
G290.1-0.8. The total molecular mass within the 95\% confidence
radius of the 3EG source is 7.7 $\times 10^5 M_\odot$ and most of
it is localized in Cloud B (4.5 $\times 10^5 M_\odot$).

Assuming typical values for the energy of the explosion
($E_{51}$=1) and the unshocked ambient density ($n=0.1 $cm$^{-3}$)
we obtain a CR enhancement factor of $\sim250$. Assuming that the
same CR enhancement is applicable to the cloud overpredicts the
EGRET flux by about a factor of 10. Even with an average CR
enhancement factor within the cloud ten times lower than within
the SNR, it is at least possible, then, that 3EG J1102-6103 and
SNR G290.1-0.8 are indeed related. Note that Bremsstrahlung, which
we have neglected, will contribute still more to the predicted
flux from SNR-cloud interactions. If the outlined scenario is
correct, GLAST and AGILE will observe a strong, compact
$\gamma$-ray source coincident with the position of Cloud B.

\subsection{ 3EG J1639-4702 -- SNR
G337.8-0.1/338.1+0.4/338.3+0.0}

Based on HI absorption seen all the way up to the terminal
velocity, Caswell et al. (1975) placed the SNR G337.8-0.1 beyond
the tangent point at 7.9 kpc. Koralesky et al. (1998) detected
maser emission in the SNR at $-45$ km s$^{-1}$, implying a far
kinematic distance of 12.4 kpc. As Figure 17 of Torres et al.
(2002) shows, there is a very massive giant molecular cloud
adjacent to the SNR in direction and close to the associated maser
in velocity ($-56$ km s$^{-1}$). The far kinematic distance for
this giant molecular cloud  is favored by (1) its likely
association with both the
  far-side maser just mentioned and a group of far-side HII regions
(group 5 in Georgelin \& Georgelin 1976); (2) its location very
close to the Galactic plane; and (3) the radius linewidth relation
for giant molecular clouds (Dame et al. 1986). The mean velocity
of the complex is $-56$ km s$^{-1}$, implying a far kinematic
distance of 11.8 kpc.

Taking the total CO luminosity of the giant molecular cloud to be
that in the range $l = 337.625$ to $338.25$, $b = -0.25$ to
$0.25$, and $v = -65$ to $-45$ km s$^{-1}$, the total molecular
mass is $5 \times 10^6 M_\odot$; this mass may be overestimated by
10-20\% owing to the inclusion of emission from gas at the same
velocity at the near kinematic distance. Even with this
correction, this giant molecular cloud ranks among the few most
massive in the Galaxy (see, e.g., Dame et al. 1986); its composite
CO linewidth of $\sim 20$ km s$^{-1}$ is correspondingly very
large. Adopting a mean radius of 0.31 deg, or 65 pc at 11.8 kpc,
the mean nucleon density of the cloud  is 176 cm$^{-3}$. The total
mass within the 95\% confidence radius of the 3EG source is $7.6
\times 10^6 M_\odot$; this mass too may be overestimated by
10-20\% owing to inclusion of near-side emission.

Again, there is so much molecular material that a conservative
enhancement factor would be enough to produce much of the GeV flux
detected by EGRET. It is at least possible that 3EG J1639-4702 is
partially related to the SNRs with which it coincides. In
addition, the Parkes pulsar PSR J1637-4642 seems to be a plausible
candidate for the origin of this 3EG source as well: Only a 12\%
efficiency would be needed (Torres, Butt, \& Camilo 2001).
Although the spectral index of the 3EG source,
$\Gamma=2.50\pm0.18$, seems quite soft in comparison with detected
EGRET pulsars, the work of Halpern et al. (2001) suggests that a
soft spectral index does not automatically rule out a pulsar
origin of the $\gamma$-rays.


\subsection{ 3EG J1903+0550 -- SNR G39.2-0.3}

Caswell et al. (1975) detected HI absorption all the way up to the
terminal velocity towards the SNR G39.2-0.3, with almost
continuous strong absorption between 60 km s$^{-1}$ and the
terminal velocity. They therefore concluded that the remnant was
certainly beyond the tangent point, and most likely at the far
distance corresponding to 60 km s$^{-1}$, $\sim$9.6 kpc. Such a
large distance is consistent with the high foreground hydrogen
column inferred both by Becker \& Helfand (1987) based on 21 cm
absorption measurements with the VLA, and by Harrus \& Slane
(1999) based on ASCA observations. A distance of 9.6 kpc would
place the SNR in the far Sagittarius arm, where the remnant is
nearly coincident with a massive molecular complex. The complex is
(40,59) in the catalog of Dame et al. (1986), who assigned the far
kinematic distance based on 2 associated HII regions.  The mass of
this complex is estimated to be $2.1 \times 10^6 M_\odot$. The
mass within the 95\% confidence radius of the 3EG source (dotted
circle in Figure 26 of Torres et al. 2002) is even higher, $3.4
\times 10^6 M_\odot$, because the radius also includes part of
another molecular complex at higher longitude. If part of the mass
contained in the molecular complex could serve as target material
for the relativistic particles accelerated in the SNR shock, this
3EG detection could plausibly be produced by a combination of
Bremsstrahlung and pion decay. With the mass quoted, a CR
enhancement factor of less than 10 is all that is needed to
produce the bulk of the observed $\gamma$-ray emission. However,
it is clear that not all of the molecular mass can be illuminated
by the SNR shock front. The SNR itself is less than 8 arcmin in
size, while the 3EG source is $\sim $1 deg in size. Only 0.1\% of
the molecular material need to serve as a target for the particles
accelerated in G39.2-0.3 in order to produce the 3EG source.

\subsection{ 3EG J2016+3657 -- SNR G74.9+1.2 (CTB 87)}

Although it has been suggested that SNR G74.9+1.2 (CTB 87) may be
interacting   with molecular clouds (Huang et al. 1983, Huang and
Thaddeus 1986), the coincident 3EG J2016+3657 source has been
proposed as a counterpart of the blazar-like radio source
G74.87+1.22 (B2013+370) (Halpern et al. 2001a, Mukherjee et al.
2000). B2013+370 is a compact, flat spectrum, 2 Jy radio source at
1 GHz. Its multiwavelength properties were compiled by Mukherjee
et al. (2000), and since they resemble other blazars detected by
EGRET, B2013+370 is an interesting possible counterpart for this
3EG source.

The Crab-like supernova remnant CTB 87 is located at more than 10
kpc (Green 2000), seemingly disfavoring its shell interactions as
the cause of the EGRET source. There are also WR stars in the
field (Romero et al. 1999), which might produce $\gamma$-ray
emission. This possibility remains to be analyzed. INTEGRAL
observations would help in determining if there is $\gamma$-ray
emission coming from the stars. Figure 27 of Torres et al. (2002)
shows a CO map for the CTB 87 region.  One clearly defined
molecular cloud appears in the map. The mean velocity of the
molecular cloud is -57 km/s. Assuming a flat rotation curve beyond
the solar circle, the cloud's kinematic distance is 10.4 kpc,
similar to that of the SNR. The total molecular mass within the
95\% confidence radius of the 3EG source is $1.7 \times 10^5
M_\odot$. With such a high value for the molecular mass, there is
still a chance that the hadronic or leptonic $\gamma$-ray emission
may be contributing to 3EG J2016+3657.


\section{CONCLUDING REMARKS}

It is at least plausible that EGRET has detected distant (more
than 6 kpc) SNRs. There are 5 coinciding pairs of 3EG sources and
SNRs for which the latter apparently lie at such high values of
distance (disregarding those related with SNRs spatially close to
the galactic center). For all these cases, we have uncovered the
existence of nearby, large, in some cases giant, molecular clouds
that could enhance the GeV signal through pion decay. It is
possible that the physical relationship between the 3EG source and
the coinciding SNR could provide for these pairs a substantial
part of the GeV emission observed. This does not preclude,
however, composite origins for the total amount of the radiation
detected. Some of these cases present other plausible scenarios
(see Torres et al. 2002 for details). AGILE observations, in
advance of GLAST, would greatly elucidate the origin for these 3EG
sources, since even a factor of 2 improvement in resolution would
be enough to favor or reject the SNR connection.

\section*{ACKNOWLEDGEMENTS}

The work of DFT was performed under the auspices of the U.S.
Department of Energy by University of California LLNL under
contract No. W-7405-Eng-48. GER and JAC. were supported by CONICET
(under grant PIP N$^o$ 0430/98), ANPCT (PICT 03-04881), as well as
by Fundaci\'on Antorchas. YMB acknowledges the support of the High
Energy Astrophysics division at the CfA and the {\em Chandra\/}
project through NASA contract NAS8-39073.

\end{document}